\documentstyle[preprint,prl,aps]{revtex}
\begin{document}
\draft

\preprint{\vbox{\hbox{March 1996}\hbox{rev. Jan. 1997}\hbox{IFP-724-UNC}}}


\title{Quarks, Squarks and Textures.}
\author{\bf Paul H. Frampton and Otto C. W. Kong}
\address{Institute of Field Physics, Department of Physics and Astronomy,\\
University of North Carolina, Chapel Hill, NC  27599-3255}
\maketitle

\begin{abstract}
By studying symmetric 
mass textures for the up and down quark sectors, and expanding in a 
small parameter $\lambda \sim sin\theta_C$, bounds are set on entries 
commonly assumed to vanish. 
Consequences of a 2 + 1 family structure which can result from 
horizontal symmetry are examined. 
Generalizing to squarks, we study 
suppression of Flavor Changing Neutral Currents by mass 
degeneracy and/or small mixing angles.

\end{abstract} 


\newpage 

\section{INTRODUCTION}
In the Standard Model, the Yukawa sector contains the majority of the
free parameters.  Though we have now a reasonably good knowledge of the
experimental values of these parameters, the flavor theory that may 
explain the family structure of the fermions and values of the parameters 
is still a major puzzle. In particular, the strong hierarchy among quark
masses and the very nontrivial structure of the CKM matrix remains a
conundrum. At a pure phenomenological level, various mass matrix {\it 
Ans\"{a}tze} have been proposed. The most popular Fritzsch  {\it Ansatz}\cite{F} had to be abandoned as we realized that the top-quark mass
is above $90GeV$\cite{90}. However, a modified version\cite{mFa,mfa}
still holds promise. The Fritzch {\it Ansatz} uses a mass matrix of
three symmetric texture zeros for both the up- and down-mass matrices, 
while the modified-Fritzch {\it Ansatz} uses one with two. In the latter
case, the zeros are at the $11-$ and $13-(31-)$entries. Having 
small entries at the locations has been shown to yield favorable relations
among the mass and mixing parameters\cite{HR}.  
Recently, the authors of  Ref~.\cite{RRR} analyzed all possible symmetric
quark mass matrices with the maximal number of texture zeros. They 
started by assigning to the up (or down) quark mass matrix any of the 
6 possible forms of symmetric matrix with an hierarchy of three non-zero
eigenvalues and three texture zeros;  they then examined admissible 
solutions with maximal number of texture zeros in the corresponding 
down (or up) quark mass matrix by fitting the experimental quarks 
masses and CKM mixing parameters RG-evolved to the GUT scale. Five
solutions are listed  (we will refer to them as RRR-textures), 
with five texture zeros and a hierarchical form with matrix entries
expressed as leading-order powers of 
$\lambda = sin\theta_C \sim 0.22$\cite{lam}. 

On the other hand, interest in the use of horizontal symmetry to derive a
phenomenologically viable quark mass texture has been resurrected 
recently\cite{u1,u1a,LNS,PS,KS,q2n,q2n2,552,725}, partly motivated 
by the possibility of obtaining
simultaneously  appropiately constrained squark (soft) mass(-squared) matrices, 
in the place of an assumed universality condition, for satisfying  the
relevant FCNC constraints\cite{sd,qsa}.
A $SU(2)$ (or $U(2)$)
horizontal symmetry with the lighter two families forming a 
doublet has then been advocated by some authors\cite{PS,q2n,552,725,sd,PT,HM,BDH}. 
This $2+1$ family structure has the favorable feature that squark degeneracy among 
the two families is guaranteed before the breaking of the horizontal symetry.
In addition, if this symmetry is  gauged, as is desirable, but its breaking goes through a
discrete subgroup, possibly dangerous $D$-term contributions which may 
lift the squark degeneracy can be avoided and it is possible 
that a model can be built with all the relevant FCNC constraints
satisfied.

Motivated by our SUSY-GUT compatible horizontal symmetry 
model building\cite{552,725}, we will  address here some 
features of the quark and squark mass matrices. In the first part, we 
derive some more general quark mass textures, using a simple algebraic
analysis. New texture patterns obtained contain less texture zeros and are
therefore less predictive. This is necessary, however,
because  the RRR-textures are in general 
incompatible with vertical unification with a $2+1$ {\it ansatz}. 
In the second part, we  look
at the squark mass  matrices and the FCNC contraints from neutral meson 
mixing, under the perspective of a $2+1$ family structure. Our analysis 
here is not constrained to GUT-scale. RG-runnings of quark mass ratios
are in general small and not important in texture pattern analysis.
For the squark masses analysis, we actually
consider low energy FCNC constraints. Some discussions and 
references on the issue is given in our 
horizontal symmetry model presentation in Ref.~\cite{725}.

In the quark mass matrix texture analysis below, our objective is
to use a simple algebraic analysis to illustrate how the textures 
are constrained by the mass and CKM parameters, and to derive some more
general texture patterns. The analysis allows us to arrive at a wide variety
of acceptable texture solutions which are phenomenologically viable.
Our solutions are compatible and include as special cases all the 
five-zero RRR-texture solutions and serve as natural generalizations
of them. This leads us to believe that we do not miss any interesting
texture pattern.  However, no attempt has been made to prove
that the textures  derived here are the only possible ones under the 
assumptions, nor do we make such a claim. 

\section{QUARK MASS MATRIX TEXTURES}
We consider a symmetric hierarchical mass matrix given as
\begin{equation}
$$M = \left(\begin{array}{ccc} 
0 & x & y \\
x & a & c \\
y & c & 1
\end{array}\right)$$ 
\end{equation} 
where $a, c, x$ and $y \leq \lambda$ (of order $\lambda^{n}, n\geq 1$). 
We have thus assumed only one zero in each mass matrix. The entry $M_{11}$
is the one that is most commonly believed to be small. Assuming it is zero
here simpifies our analysis. We will later show limits on the entry 
that can be admitted without upsetting our texture pattern solutions.
Starting with the only order one entry $M_{33}$ , we put in the small entries
 and obtain the three eigenvalues together with the diagonalizing matrix
$V$ through a perturbational approach. For all the numbers, {\em we are
 interested only in their approximate values as represented by orders in
$\lambda$}, hence we keep only potentially leading-order terms.

We put in $a$
and $x$ first, and then $c$ and $y$, the latter as a perturbation to the 
diagonalized matrix from the former. We have then
\begin{equation}
$$VMV^{\dag} = V_2 V_1  \left(\begin{array}{ccc} 
0 & x & y \\
x & a & c \\
y & c & 1
\end{array}\right)$$ V_{1}^{\dag} V_{2}^{\dag}
\sim V_2   \left(\begin{array}{ccc} 
-x^{2}/a & 0 & y - c x/a \\
0 & a + x^{2}/a & c \\
y - c x/a & c & 1
\end{array}\right)$$  V_{2}^{\dag}
\sim M(diag) \ ,
\end{equation}
 
\begin{equation}
$$V_{1}^{\dag} \sim \left(\begin{array}{ccc} 
1 & x/a & 0 \\
-x/a & 1 & 0 \\
0 & 0 & 1
\end{array}\right)$$ \ ,
\end{equation} 
and
\begin{equation}
$$V_{2}^{\dag} \sim \left(\begin{array}{ccc} 
1 & 0 & y - c x/a \\
0 & 1 & c/(1-a) \\
-(y - c x/a) & -c/(1-a) & 1
\end{array}\right)$$ \ . 
\end{equation} 
Here we note that in the first step, where $x$ is treated as a perturbation,
$x$ must be taken smaller than $a$ ({\it i.~e.} $x$ is at least one order 
higher in $\lambda$).  We take this assumption here and leave 
the alternative situation to be handled later. Specifically, we assume
$x \leq \lambda ^{2} $ and $x/a \leq \lambda$.  
The final result is as follows:
\begin{equation}
$$V^{\dag} = V_{1}^{\dag}V_{2}^{\dag} \sim \left(\begin{array}{ccc} 
1 & x/a & (y - c x/a)\\
-x/a & 1 & c - y x/a \\
-(y - c x/a) & -(c - y x/a) & 1
\end{array}\right)$$ \ , 
\end{equation}
\begin{equation}
M(diag) \sim \{ \ \ -\frac{x^{2}}{a} -(y - c \frac{x}{a})^2,  \ \ 
a +\frac{x^{2}}{a} -\frac{c^2}{1-a}, \ \ 
1 +\frac{c^2}{1-a} + (y - c \frac{x}{a})^2 \ \ \} \ . 	\label{mq}
\end{equation}
Note that for the mass eigenvalue expressions, Eq.(6), we have chosen to 
display the principal terms  from each  entry to the matrix, not just
the possible overall leading order terms. For instance, 
$x^2/a$ cannot be a overall leading order term for the middle eigenvalue as
the same term contributes to the smallest eigenvalue.

The result can then be applied to both the up- and down-quark mass matrices.
We take $M^{u}(diag)$ and $V_{u}^{\dag}$ as given by the above equations 
and $M^{d}(diag)$ and $V_{d}^{\dag}$ as given by analog equations with
$a, x, c$ and $y$ replaced by $a', x', c'$ and $y'$; with the
masses normalized to $m_t = 1$ and $m_b = 1$ respectively. In addition to
the expressions for the mass eigenvalues, we have also the elements of
the CKM-matrix ($V_{CKM}=V_{u}V_{d}^{\dag}$); all these parameters can be 
derived from experimental measurements and expressed in powers of 
$\lambda$\cite{RRR,LNS}.
Assuming no delicate cancellation of numbers in the expressions of
the mass eigenvalues and mixings, at least one term in each 
must be of the right order in $\lambda$. Hence we arrive at the 
following list of constraints:
\begin{enumerate}
\item
$ x (x/a), (y - c x/a)^2 \leq \lambda^{8} $;
\item
$ a, c^2 \leq \lambda^{4} $;
\item
$ x' (x'/a'), (y' - c' x'/a')^2 \leq \lambda^{4} $;
\item
$ a', c'^{2} \leq \lambda^{2} $;
\item
$ x/a, x'/a' \leq \lambda $;
\item
$ c, c' \leq \lambda^{2} $;
\item
$  y', c' x'/a', c' x/a, c x'/a' \leq \lambda^{3} (\sim \lambda^{4})$.
\end{enumerate}
where at least one term in  each case must 
satisfy the equality, rather than inequality.
In the last three constraints, which are from the CKM-mixings,
we have left out terms whose magnitudes have an upper bound already
more strongly constrained by the  mass matrices.
 Combining the 
conditions\cite{yc} then leads to
\[ a' \sim \lambda^{2}, \ \ \ \ \  x' \sim \lambda^{3}, \ \ \ \ \ \  a \sim \lambda^{4}, \]
with the following solutions:
\begin{itemize}
\item Case 1 
\[
$$M^{u} \sim \left(\begin{array}{ccc} 
(\leq \lambda^{8}) & \lambda^{6} & \leq \lambda^{5} \\
\lambda^{6} & \lambda^{4} & \lambda^{2} \\
\leq \lambda^{5} & \lambda^{2} & 1
\end{array}\right)$$ 
\ \
OR \ \
$$M^{u} \sim \left(\begin{array}{ccc} 
(\leq \lambda^{8}) & \leq \lambda^{6} & \lambda^{4} \\
\leq \lambda^{6} & \lambda^{4} &  \lambda^{2} \\
\lambda^{4} &  \lambda^{2} & 1
\end{array}\right)$$,  \ \ \ \ 
$$M^{d} \sim \left(\begin{array}{ccc} 
(\leq \lambda^{4}) & \lambda^{3} & \leq \lambda^{3} \\
\lambda^{3} & \lambda^{2} & \leq \lambda^{2} \\
\leq \lambda^{3} & \leq \lambda^{2} & 1
\end{array}\right)$$ ;
\]
\item Case 2
\[
$$M^{u} \sim \left(\begin{array}{ccc} 
(\leq \lambda^{8}) & \lambda^{6} & \leq \lambda^{4} \\
\lambda^{6} & \lambda^{4} & \leq \lambda^{2} \\
\leq \lambda^{4} & \leq \lambda^{2} & 1
\end{array}\right)$$ 
\ \ 
OR \ \
$$M^{u} \sim \left(\begin{array}{ccc} 
(\leq \lambda^{8}) & \leq \lambda^{6} & \lambda^{4} \\
\leq \lambda^{6} & \lambda^{4} & \leq \lambda^{2} \\
\lambda^{4} & \leq \lambda^{2} & 1
\end{array}\right)$$, 
\ \ \ \ 
$$M^{d} \sim \left(\begin{array}{ccc} 
(\leq \lambda^{4}) & \lambda^{3} & \leq \lambda^{3} \\
\lambda^{3} & \lambda^{2} & \lambda^{2} \\
\leq \lambda^{3} & \lambda^{2} & 1
\end{array}\right)$$. 
\]
\end{itemize}
Note that the $M_{11}$ entries are limits, put in  {\it a posteriori},
that can be allowed without upsetting the solutions. Allowing any $M_{11}$ 
entry to take the its maximum value to begin with will however 
modify the constraints.

Alternatively, one can  put in $c$
and $a$ first, and then $x$ and $y$. We have then
\begin{equation}
$$VMV^{\dag} = V_2 V_1 M V_{1}^{\dag} V_{2}^{\dag} 
\sim V_2   \left(\begin{array}{ccc} 
0 & x-cy & xc+y  \\
x-cy & -c^2+a  & 0 \\
xc+y & 0 & 1
\end{array}\right)$$  V_{2}^{\dag}
\sim M(diag)
\end{equation}
 
\begin{equation}
$$V_{1}^{\dag} \sim \left(\begin{array}{ccc} 
1 & 0 & 0 \\
0 & 1 & c \\
0 & -c & 1
\end{array}\right)$$ 
\end{equation} 

\begin{equation}
$$V_{2}^{\dag} \sim \left(\begin{array}{ccc} 
1 & -(x-cy)/(c^2-a) & (xc+y)/(1+c^2)  \\
(x-cy)/(c^2-a) & 1 & 0 \\
-(xc+y)/(1+c^2) & 0 & 1
\end{array}\right)$$ 
\end{equation}

and
\begin{equation}
M(diag) \sim  \{ \ \ \frac{(x-cy)^2}{(c^2-a)}  -\frac{(xc+y)^2}{(1+c^2)}, \ \ 
a -c^2 -\frac{(x-cy)^2}{(c^2-a)}, \ \ 
1 +c^2 + \frac{(xc+y)^2}{(1+c^2)} \ \ \} .
\end{equation}
Here we aim at alternative texture patterns that need not satisfy $x/a \leq \lambda$.
Hence, we consider only the case with $a << c^2$ (at least one higher order in $\lambda$). Note that we need also
$x/c^2, y/c \leq \lambda$ for the perturbational approximation to be valid.  Taking these expressions for $M^{u}(diag)$ and $V_{u}^{\dag}$ and the previous
result for the down-sector to  repeat the analysis, we obtain an alternative
texture patterns as  
\begin{itemize}
\item Case 3
\[
$$M^{u} \sim \left(\begin{array}{ccc} 
(\leq \lambda^{8}) & \lambda^{6} & \leq \lambda^{4} \\
\lambda^{6} & \leq \lambda^{5} & \lambda^{2} \\
\leq \lambda^{4} & \lambda^{2} & 1
\end{array}\right)$$ 
\ \ \ \ \ \ 
$$M^{d} \sim \left(\begin{array}{ccc} 
(\leq \lambda^{4}) & \lambda^{3} &  \leq \lambda^{3} \\
\lambda^{3} & \lambda^{2} & \leq \lambda^{2} \\
\leq \lambda^{3} & \leq \lambda^{2} & 1
\end{array}\right)$$. 
\]
\end{itemize}
Compared with the previous result, one can see easily that taking 
this alternative approach to $M^{d}$ ({\it i.~e.} $a' << c'^{2}$) does
not lead to any consistent solution.

Starting from  simple assumptions, therefore, we have succeeded in deriving the 
above hierarchical mass texture patterns\cite{PW}.

Here we compare our result with the RRR-textures. Unlikely, the latter, we
have not go through a detailed numerical analysis. However, we start with 
a much more general form for the mass matrices and show that the simple
algebraic analysis is powerful enough for us to obtain the various 
texture patterns, with undetermined coefficients of order unity. 
The texture patterns are in a sense generalizations of
the RRR-textures. Note that in the latter analysis the zeros in general
do not have to be exact. 
For instance, replacing a zero with an entry higher order in $\lambda$ 
than all the other entries in the matrix will not upset the solution.
We see that without the prior assumption of the
existence of many texture zeros, some of the
small entries in the mass matrices can actually be much larger than one
would expect them to be, from naively applying the RRR-textures.
This would be of interest from the model building perspective.

For a detail comparison, first we note that our result gives a down-quark
mass matrix always in the form
\begin{equation}
$$M^{d} \sim \left(\begin{array}{ccc} 
* & \lambda^{3} &  * \\
\lambda^{3} & \lambda^{2} & * \\

* & * & 1
\end{array}\right)$$ \ \ ; 
\end{equation}
while the common structure for all the RRR-textures has the form
\begin{equation}
$$M^{d} \sim \left(\begin{array}{ccc} 
* & \lambda^{4} &  * \\
\lambda^{4} & \lambda^{3} & * \\

* & * & 1
\end{array}\right)$$ \ \ . 
\end{equation}
A power of $\lambda$ analysis of the latter form gives
\begin{equation}
M^{d}(diag) \sim  \{\ \  \lambda^{5}, \ \ \ \ \lambda^{3}, \ \ \ \ 1 \ \ \}
\end{equation}
instead of the more popular 
\begin{equation}
M^{d}(diag) \sim  \{\ \  \lambda^{4}, \ \ \ \ \lambda^{2}, \ \ \ \ 1 \ \ \} 
\end{equation}
that we used.  It can be checked easily that if we started by putting
the former $M^{d}(diag)$ into  constraints 3 and 4 in our list, all of our analysis
would go through with the only modification in our solutions given by
changing  $M^{d}$ to the form of Eq.~(11).
Recall that $\lambda \sim 0.22$, and order one coefficients are allowed
in all the terms. Large coefficients would easily change the order of
$\lambda$ result.  This kind of ambiguity is unfortunately unavoidable
in the type of order in $\lambda$ analysis.  The approach is a useful one
 for obtaining mass matrix 
{\it ans\"{a}tze} or textures\cite{lamb} but not exact results. 
To further our comparison, we will assume
this alternative $M^{d}$ solutions for all three cases. Then, the only apparent
conflict of our $M^{d}$ results  with the RRR-textures is that the $23/32$-entry
in case 2 is fixed at $\lambda^{2}$, while the correspondent entry in RRR-textures, if 
not zero, is given by $\lambda^{3}$. However, there is a large coefficient
of $4$ from their numerical analysis. 
Hence a coefficient a bit smaller than one for our case would reconcile the 
difference. A conflict appears in the limiting form of  case 3, 
which indicates that a six-zero
texture pattern is admitted by the naive algebraic analysis, while six-zero cases are ruled out in the 
RRR-analysis. This particular six-zero pattern remains actually a very popular
candidate\cite{60}. Putting in a $\lambda^{3}$ term for the $23/32$-entry of $M^{d}$ while keeping the other
 zeros, however,  does give one of the
RRR-textures. Otherwise, the other RRR-textures all fit in with our patterns. 
Detailed numerical analysis is of course useful to further establishing the
viablity of the texture patterns obtained here. Nevertheless, so far as a 
texture pattern with less zeros is concerned, the extra
coefficients definitely give more flexiblility for fitting the experimental
parameters. Hence  we do expect these texture patterns to be valid, except possibly 
the six-zero texture.

\section{SQUARK MASS MATRICES AND FCNC CONSTRAINTS}
Now we turn to the scalar quark sector and look into how a $2+1$ family
structure fits into the squark mediated FCNC contraints from neutral meson mixings.
First note that the  squark mass  matrices $\tilde{M}^{u2}$ and
$\tilde{M}^{d2}$ are each divided into four $3\times 3$ sub-matrices as
\begin{equation}
$$\tilde{M}^{u2} = \left(\begin{array}{cc} 
\tilde{M}^{u2}_{LL} & \tilde{M}^{u2}_{LR} \\
(\tilde{M}^{u2}_{LR})^{\dag} & \tilde{M}^{u2}_{RR} \\
\end{array}\right), \ \ \ \ 
\tilde{M}^{d2} = \left(\begin{array}{cc} 
\tilde{M}^{d2}_{LL} & \tilde{M}^{d2}_{LR} \\
(\tilde{M}^{d2}_{LR})^{\dag} & \tilde{M}^{d2}_{RR} \\
\end{array}\right)$$ . 
\end{equation}
The leading contributions to the off-diagonal blocks arise from
the trilinear $A$-terms, while the leading contributions to the diagonal blocks
arise from the soft mass terms. The latter dominate over the former,
and can generally lead to unacceptably large FCNC-effect in neutral meson mixing
when universality of soft masses is not imposed. Hence they are our
subject of concern here\cite{Afc}.
We start by considering the following (diagonal block) mass  matrix for 
general squarks
\begin{equation}
$$\tilde{M}^2 = \tilde{m}^2 \left(\begin{array}{ccc} 
\tilde{a}& \tilde{x} & (\tilde{c} + \tilde{y} )/\sqrt{2} \\
\tilde{x} & \tilde{a}& (\tilde{c} - \tilde{y} )/\sqrt{2} \\
(\tilde{c} + \tilde{y} )/\sqrt{2} & (\tilde{c} - \tilde{y} )/\sqrt{2} & \tilde{b}
\end{array}\right)$$ 
\end{equation} 
where $\tilde{a}, \tilde{b}$ are order $1$ and $\tilde{c} , \tilde{x} , \tilde{y} \leq \lambda$.
Note that order $1$ quantities are expected for the diagonal entries as the
correspondent mass terms are naturally invariant under any horizontal symmetry.
Equality of the first two diagonal entries is dictated by the $2+1$ structure
which we are interested in here. The symmetry structure also suggests that any
higher dimensional, horizontal symmetry breaking, mass term naturally 
gives the same contributions to the two entries, and hence lifting in degeneracy of the two mass eigenvalues can be attributed to only the contributions of
the non-diagonal terms\cite{aa}.

We first take a rotation to diagonalize the upper two by two block and add in 
the rest by a perturbational analysis: 
\begin{equation}
$$\tilde{V} \tilde{M}^2 \tilde{V}^{\dag} = 
\tilde{V}_0 V_0  \tilde{M}^2 V_{0}^{\dag} \tilde{V}_{0}^{\dag} 
\sim \tilde{V}_0  \left(\begin{array}{ccc} 
\tilde{a}-\tilde{x} & 0 & \tilde{y}  \\
0 & \tilde{a}+ \tilde{x}  & \tilde{c} \\
\tilde{y} & \tilde{c} & \tilde{b}
\end{array}\right)$$ \tilde{V}_{0}^{\dag} 
\sim \tilde{M}^2(diag)
\end{equation}
with 
\begin{equation}
$$V_{0}^{\dag} = \left(\begin{array}{ccc} 
1/\sqrt{2} & 1/\sqrt{2} & 0 \\
-1/\sqrt{2} & 1/\sqrt{2} & 0 \\
0 & 0 & 1
\end{array}\right)$$;  \label{v0}
\end{equation} 
then
\begin{equation}
$$\tilde{V}_{0}^{\dag} \sim \left(\begin{array}{ccc} 
1 & 0 & \tilde{y} /(\tilde{b} -\tilde{a}+ \tilde{x} )   \\
0 & 1 & \tilde{c} /(\tilde{b} -\tilde{a}- \tilde{x} )   \\
-\tilde{y} /(\tilde{b} -\tilde{a}+ \tilde{x} ) & -\tilde{c} /(\tilde{b} -\tilde{a}- \tilde{x} ) & 1
\end{array}\right)$$ 
\end{equation}
and
\begin{eqnarray}
\tilde{M}^2(diag) \sim  & \tilde{m}^2 \{ \ \ \tilde{a}-\tilde{x} & -\frac{\tilde{y} ^{2}}{\tilde{b} - \tilde{a}+ \tilde{x}}, \ \ 
\tilde{a}+\tilde{x} -\frac{\tilde{c} ^{2}}{\tilde{b} - \tilde{a}- \tilde{x} }, \nonumber \\
 &  &  \tilde{b} + \frac{\tilde{c} ^{2}}{\tilde{b} - \tilde{a}- \tilde{x} } + \frac{\tilde{y} ^{2}}{\tilde{b} - \tilde{a}+ \tilde{x}} \ \ \}.
\end{eqnarray}
Note that the difference $(\sim \tilde{b} -\tilde{a})$ between the third and 
the first or second eigenvalues,  is of order one (in $\tilde{m}^2$), 
while the degeneracy between the first and second eigenvalues is lifted by
\begin{equation}
\Delta \tilde{m}^2_{12} \sim 2\tilde{x} -\tilde{c} ^{2} + \tilde{y} ^{2}
\end{equation}

The other quantity that affects the FCNC is the squark mass mixing 
matrix which, in the diagonal quark mass basis, is generally expressed as 
\begin{equation}
K = V \tilde{V}^{\dag} .
\end{equation} 
To be specific, this includes
\begin{eqnarray}
K_{L}^{d} = V_{L}^{d} \tilde{V}_{L}^{d\dag}, \ \ \ \ 
K_{R}^{d} = V_{R}^{d} \tilde{V}_{R}^{d\dag},  \nonumber \\
K_{L}^{u} = V_{L}^{u} \tilde{V}_{L}^{u\dag}, \ \ \ \
K_{R}^{u} = V_{R}^{u} \tilde{V}_{R}^{u\dag};
\end{eqnarray}
and the $V$'s and $\tilde{V}$'s are diagonalizing matices for quarks.
\begin{equation}
V_{L}^{d} M^{d} V_{R}^{d\dag} = M^{d}(diag), \ \ \ \ 
V_{L}^{u} M^{u} V_{R}^{u\dag} = M^{u}(diag), \ \ \ \
\end{equation}
and for squarks,
\begin{eqnarray}
\tilde{V}_{L}^{d} \tilde{M}^{d2}_{LL} \tilde{V}_{L}^{d\dag} = \tilde{M}^{d2}_{LL}(diag),
\tilde{V}_{R}^{d} \tilde{M}^{d2}_{RR} \tilde{V}_{R}^{d\dag} = \tilde{M}^{d2}_{RR}(diag), \nonumber \\
\tilde{V}_{L}^{u} \tilde{M}^{u2}_{LL} \tilde{V}_{L}^{u\dag} = \tilde{M}^{u2}_{LL}(diag),
\tilde{V}_{R}^{u} \tilde{M}^{u2}_{RR} \tilde{V}_{R}^{u\dag} = \tilde{M}^{u2}_{RR}(diag).
\end{eqnarray}
We will however suppress subscripts and superscripts wherever unambiguious.

Taking the hierarchical form of $M$, as for example given by one of
our texture pattern solutions, we have
\begin{equation}
K = (V_{2}V_{1}) V_{0}^{\dag} \tilde{V}_{0}^{\dag}
\sim $$\left(\begin{array}{ccc} 
1/\sqrt{2} & 1/\sqrt{2} & \tilde{c} /\sqrt{2}  -y + c x/a     \\
-1/\sqrt{2} & 1/\sqrt{2} & \tilde{c} /\sqrt{2}   - c   \\
-c/\sqrt{2}  - \tilde{y} & c/\sqrt{2}  - \tilde{c}  & 1
\end{array}\right)$$ \label{k}
\end{equation}
where we have replaced $\tilde{b} - \tilde{a}$ by $1$ and keep only the would be leading
order terms.
However, if we start with 
\begin{equation}
$$M = \left(\begin{array}{ccc} 
a/2 + x & a/2 & (c+y)/\sqrt{2} \\
a/2 & a/2 - x & (c-y)/\sqrt{2} \\
(c+y)/\sqrt{2} & (c-y)/\sqrt{2} & 1
\end{array}\right) 
= V_{0}^{\dag} \left(\begin{array}{ccc} 
0 & x & y \\
x & a & c \\
y & c & 1
\end{array}\right) V_0 $$ \ ,	\label{mqpd} 
\end{equation} 
then we have 
\begin{equation}
K = (V_{2}V_{1}) V_0 V_{0}^{\dag} \tilde{V}_{0}^{\dag}
\sim $$\left(\begin{array}{ccc} 
1 & -x/a  +cy +\tilde{c} y  & \tilde{y} -y +c x/a -\tilde{c} x/a     \\
x/a -cy + c \tilde{y}  & 1 &  \tilde{c}  -c +y x/a +\tilde{y} x/a   \\
 y -\tilde{y} -c x/a &   c -\tilde{c} -y x/a  & 1
\end{array}\right)$$. \label{kpd}
\end{equation}
Compare the two expressions, one can see that the latter case gives,
 in general, smaller mixings.

From the perspective of horizontal symmetry,
while a hierarchical quark mass matrix, rank one in
first order, can be easily enforced, squark mass  matrices
$\tilde{M}^2_{LL}$ and $\tilde{M}^2_{RR}$ naturally have 
all their  diagonal entries of order one. 
 Unless they are part of the same multiplet, equality
of the diagonal squark mass  entries does not come naturally. Our choice of diagonal
entries is dictated by $2+1$ family structure. The degeneracy is of course
lifted by the off-diagonal entries. The first observation here is that when
a degeneracy will be lifted by perturbation, the resultant eigenstates are naturally
given by a maximal mixing, and hence by the first $2\times 2$ block in 
Eqs.~(\ref{v0}) and (\ref{k}). 

In the quark-squark alignment(QSA) approach\cite{LNS,qsa}, one 
gives up the squark mass  degeneracy requirement
for FCNC suppression.  If the quark and squark mass matrices
could be almost diagonalized  simultaneously, the mixing matrix $K$ would have
small off-diagonal elements and hence give the necessary FCNC 
suppression. However, there
is also {\em no} easy way to obtain such a result 
from a horizontal symmetry. As shown above, the
degeneracy approach goes to the other extreme, in favor of maximum 
mixing, for the lighter two generations in our case.
After all, in first order form  the quark and squark mass matrices are
expected to be very different. Is there a way to reconcile this with the 
QSA? Within the $2+1$ family structure, quark mass matrices of the form given
by Eq.~(\ref{mqpd}) seem to give the answer. The first order mixing in $K_{12}$
is removed, as shown in Eq.~(\ref{kpd}).

The form  of the quark mass matrix can be described as
democratic in the first two families
and hierarchical between them and the third.
It takes only a simultaneous rotation
of $M^{u}$ and $M^{d}$ given by any of the above (phenomenologically viable)
 hierarchical texture patterns by $V_0$ to give a pair of matrices in 
this desired form. 
We consider it an interesting alternative of {\it partial} quark-squark alignment. 
In particular, in the $SU(5)$ unification or a $2+1$ family structure 
framework,   the symmetric nature of $M^{u}$ and the need to have a
relatively large $V_{us}$ make our partial alignment appear as the 
best option for suppression of $K_{12}$. The necessary suppression of $K_{23}$ 
or $K_{31}$ can be easily  obtained  even without  alignment.
Quark mass matrices in this form together
with the required squark mass matrices can be derived naturally from a $Q_{2N}$
symmetry\cite{725}.

Let us complete the analysis by  taking a look at  the FCNC constraints from the neutral meson
mixing\cite{FCNC} and how they can be satisfied within our framework. For
instance, constraints from $K-\bar{K}$ and $B-\bar{B}$ mixing on 
$\tilde{M}^{d2}_{LL}$ can be expressed by an upper bound on
\begin{equation}
(\delta ^d_{LL})_{12} = \frac{1}{\tilde{m}^2} (\tilde{m}^2 _1 K_{11} K^{\dag}_{12} +  \tilde{m}^2 _2
 K_{12}  K^{\dag}_{22} + \tilde{m}^2 _3  K_{13}  K^{\dag}_{32})
\end{equation}
and
\begin{equation}
(\delta ^d_{LL})_{13} = \frac{1}{\tilde{m}^2} (\tilde{m}^2 _1 K_{11} K^{\dag}_{13} +  \tilde{m}^2 _2
 K_{12}  K^{\dag}_{23} + \tilde{m}^2 _3  K_{13}  K^{\dag}_{33})
\end{equation}
respectively, where $\tilde{m}^2 _i$ are the three eigenvalues and $K$ is
actually $K^d_L = V^d_L\tilde{V}^{d\dag}_L$ with $\tilde{V}^{d}_L$ being
the unitary matrix that diagonalize $\tilde{M}^{d2}_{LL}$. All the numerical 
bounds of the type are shown in Table 1. With $K^d_L = V^d_L\tilde{V}^{d\dag}_L$
given by the form in Eq.~(\ref{kpd}), the above analysis ($x'/a' \sim \lambda$)  leads to
\begin{eqnarray}
(\delta ^d_{LL})_{12} & \sim \Delta \tilde{m}^2_{12} (K^d_{L})_{12} \nonumber \\
			& \sim \lambda  (2\tilde{x} -\tilde{c} ^{2} + \tilde{y} ^{2})  
\end{eqnarray}
which gives the bound  $\tilde{x}, \tilde{c} ^{2}, \tilde{y} ^{2} \leq \lambda^{3}$. The case for the $(\delta ^d_{LL})_{13}$ constraint looks more complicated.
However, if we take $\tilde{c} \leq \lambda ^{2}$ and $\tilde{y} \leq \lambda ^{3}$, we 
would have 
\begin{equation}
$$K^d_{L} \sim \left(\begin{array}{ccc} 
1 & \lambda & \leq \lambda^{3} \\
\lambda & 1 & \leq \lambda^{2} \\
\leq \lambda^{3} & \leq \lambda^{2} & 1
\end{array}\right)$$ 
\end{equation}
giving easily $(\delta ^d_{LL})_{13} \leq \lambda^{3}$, for example.

This illustrates how sufficient FCNC suppression can be obtained within this scheme. Details of the various constraints on the parameters in the squark
mass matrix of the form given by Eq. (16) is listed in Table 2.

For an explicit application of the kind of algebraic {\it ansatz} presented here, 
readers are referred to our $Q_{12}\otimes U(1)$ horizontal symmetry model
built along the pattern\cite{725}.

\acknowledgements
This work was supported in part by the U.S. Department of 
Energy under Grant DE-FG05-85ER-40219, Task B.\\

\bigskip
\bigskip

{\bf Table Caption.}\\

Table 1: FCNC constraints from neutral meson mixings.
The numerical bounds are given as an illustrative set of values (from
Ref.~\cite{qsa}), details of  which depend on gaugino and squark masses.
Necessary suppressions in powers of $\lambda$ are also given.

\bigskip

Table 2: Details of the constraints on our squark mass matrix parameters.
A special point to note is that we impose only the 
 $(\delta _{LL})$ 	and  $(\delta _{RR})$ constraints from Table 1,
but not the  $\left \langle \delta \right \rangle$ ones. For the case of
 $\left \langle \delta ^d_{12} \right \rangle$, it actually leads to a stronger
constraint. QSA stands for quark-squark alignment; partial QSA as described
in the text has quark mass matrices of the form given by Eq. (27).

\newpage

\begin{tabular}{c|cccc}\hline\hline
$K-\bar{K}$ mixing	& $(\delta ^d_{LL})_{12}$ 	& $(\delta ^d_{RR})_{12}$	& $\left \langle \delta ^{d}_{12}\right \rangle$\\ \hline
upper bound		& $0.05$ & $0.05$ & $0.006$\\ \hline
			& $\lambda ^{3}$ & $\lambda ^{3}$ & $\lambda ^{4}$	\\ \hline \hline
$B-\bar{B}$ mixing	& $(\delta ^d_{LL})_{13}$	& $(\delta ^d_{RR})_{13}$	& $\left \langle \delta ^{d}_{13}\right \rangle$\\ \hline
upper bound		& $0.1$  & $0.1$  & $0.04$  \\ \hline
			& $\lambda ^{2}$ &  $\lambda ^{2}$ & $\lambda ^{2}$	\\ \hline \hline
$D-\bar{D}$ mixing	& $(\delta ^u_{LL})_{12}$	& $(\delta ^u_{RR})_{12}$	& $\left \langle \delta ^{u}_{12}\right \rangle$\\ \hline
upper bound		& $0.1$  & $0.1$  & $0.04$   \\ \hline
			& $\lambda ^{2}$ & $\lambda ^{2}$ & $\lambda ^{2}$ 	\\ \hline \hline
\end{tabular}

\vspace{.5in}

Table 1: FCNC constraints from neutral meson mixings.
The numerical bounds are given as an illustrative set of values (from
Ref.~\cite{qsa}), details of  which depend on gaugino and squark masses.
Necessary suppressions in powers of $\lambda$ are also given.

\vspace*{1in}

\begin{tabular}{c|ccc|ccc}\hline\hline
  & \multicolumn{3}{c|}{with  partial QSA} &
 \multicolumn{3}{c}{without  partial QSA}\\ \hline
$\tilde{M}^{d2}_{LL} \ , \tilde{M}^{d2}_{RR}$  & $\tilde{x} \leq \lambda^3$ 
& $\tilde{c} \leq \lambda^2$
& $\tilde{y} \leq \lambda^2$ & $\tilde{x} \leq \lambda^4$
  &  $\tilde{c} \leq \lambda^2$ &   $\tilde{y} \leq \lambda^2$ \\ \hline  
$\tilde{M}^{u2}_{LL} \ , \tilde{M}^{u2}_{RR}$ 
  & $\tilde{x} \leq \lambda$ &   $\tilde{c^2} \leq \lambda$
 & $\tilde{y^2} \leq \lambda$ &
$\tilde{x} \leq \lambda^2$  & $\tilde{c} \leq \lambda$ 
  & $\tilde{y} \leq \lambda$ \\ \hline\hline
\end{tabular}

\vspace{0.5in}

Table 2: Details of the constraints on our squark mass matrix parameters.
A special point to note is that we impose only the 
 $(\delta _{LL})$ 	and  $(\delta _{RR})$ constraints from Table 1,
but not the  $\left \langle \delta \right \rangle$ ones. For the case of
 $\left \langle \delta ^d_{12} \right \rangle$, it actually leads to a stronger
constraint. QSA stands for quark-squark alignment; partial QSA as described
in the text has quark mass matrices of the form given by Eq. (27).


\begin{references}
\bibitem{yc}
Notice that the first two constraints, can lead to an
apparent solution which has both the two lower eigenvalues ($m_u$ and $m_c$)
generated  just from the mixing with the third family.
This actually gives a vanishing determinant for the mass matrix.
For instance in   Eq.~(\ref{mq}, the case with $y^2$ and $c^2$ as the leading
contributions to $m_u$ and $m_c$ respectively. This admits
a solution of the form
\[ M \sim \left( \begin{array}{ccc}
0 & 0 & y\\
0 & 0 & c\\
y & c & 1
\end{array}\right) \]
 which has zero determinant!
This "solution" of course has to be discarded.
\bibitem{PW}
During the preparation of the manuscript, the authors came 
across Ref.~\cite{PW2} which analyzes general texture patterns at low-energy,
 and their RG-evolution to GUT scale. Their "natural mass matrix"
 philosophy is similar to that of our approach. The actual analysis 
 and solutions are not the same. Among other things, our texture
 patterns, with some entries given as inequalities, give more general
 "natural" patterns.
\bibitem{Afc}
We neglect  in our analysis FCNC constraints from $\tilde{M}^2_{LR}$;
assuming proportionality  to the Yukawa couplings, such contraints can be
easily satisfied.
\end{references}
\end{document}